\documentclass[%
 reprint,
nofootinbib,
 amsmath,amssymb,
 aps,
]{revtex4}

\usepackage{graphicx}
\usepackage{dcolumn}
\usepackage{bm}
\usepackage{mathrsfs}
\usepackage{bbold}

\def\beq{\begin{eqnarray}}
\def\eea{\end{eqnarray}}
\def\be{\begin{equation}}
\def\ee{\end{equation}}

\begin{document}

\title{Superstring entanglement at finite temperature and its Hagedorn behavior }

\author{Daniel Luiz Nedel}
 \affiliation{Universidade Federal da Integra\c{c}\~{a}o Latino-Americana, Instituto Latino-Americano de Ci\^{e}ncias da Vida e da Natureza, Av. Tancredo Neves 6731 bloco 06, CEP: 85867-970, Foz do Igua\c{c}u, PR, Brasil}
 \email{daniel.nedel@unila.edu.br}





\begin{abstract}


This work demonstrates that a superstring extended density matrix can be defined at finite temperature. This enables the calculation of  the extended entanglement entropy between string coordinates. Using a real-time approach, the entanglement entropy is expressed  in terms of modular functions, whose properties are then used to study Hagedorn behavior.
\end{abstract}
\maketitle 

\newpage
\section{Introduction.}
Entanglement is a fundamental feature of quantum
theories.  The measure of entanglement, the Entanglement Entropy (EE),  has been playing an important role in several areas of theoretical physics. For example: in condensed matter, the entanglement provides a useful order parameter in topological theories
\cite{Fendley:2006gr},\cite{PhysRevLett.96.110405}, \cite{Levin:2006zz}; in conformal theories the EE is used to detect central charges \cite{Holzhey:1994we}, \cite{Calabrese:2004eu}; in the AdS / CFT context the EE helps to understand how to store quantum data in geometric objects, such as the area of a given surface \cite{RT}.

So, given the importance of entanglement entropy in quantum physics, if we have a way to bipartite a system C of the form $C= A\bigcup B$, and given a state, it is crucial to ask how entangled A and B are. An important way to bipartite a quantum system is to separate the degrees of freedom living in different spatial subregions. In general, in quantum field theories, the EE for spatial subregions is computed using the replica trick \cite{Rangamani_2017}. In this case, the EE is not computed directly. Rather, the Rényi entropy is computed and the entanglement entropy itself is obtained via an analytic continuation. For 2d conformal theories, an algebraic method has been proposed to compute the EE directly in \cite {Dias:2020srq}.  

In string theory, it is difficult to define the EE for spatial subregions
due to the extended nature of strings. This problem was addressed in references 
 \cite{He:2014gva}, \cite {Mertens:2015adr}, where the replica trick was used to compute the one-loop 
 entanglement entropy across a Rindler horizon. In \cite{ Balasubramanian:2018axm},  the subregion was chosen inside a Cauchy surface
in the space of open string configurations, and the light cone string field theory was used to compute the open string EE for spatial subregions. Unlike particle physics, these results show that the EE in string theory is finite in the UV limit. This is just because the $\alpha^{\prime}$ parameter works as a natural cutoff.

Although the EE associated with spatially separated degrees of freedom has attracted much interest, there are many ways to bipartite the Hilbert space in quantum field theories. For example, in \cite{Balasu} it was demonstrated that the EE between subsets of degrees of freedom at different momentum scales is a natural observable and a way to calculate the momentum space entropy was presented. Concerning closed strings, boundary states associate to D-branes are in fact a left/right entanglement state and it was shown in \cite{z2} that the left/right
entanglement entropy provides a suitable generalization of boundary entropy and of the
D-brane tension.  The left/right  entanglement also appears when the closed string propagates in time dependent pp wave backgrounds (which admit an interpretation in terms of null cosmological geometries) \cite{ GMN}, \cite{Marchioro:2020qub} . In this case the time dependent geometry dynamically generates a left/right EE and the vacuum state, as seen by asymptotic observers, is a time dependent left/right entangled state. For the background presented in \cite{PRT}, the EE diverges near the singularity and the vacuum at the singularity is a D-brane state \cite{Marchioro:2007ch}. On the other hand, for the background presented in \cite{Bin} it was shown in \cite{Marchioro:2020qub}  that near the singularity the left/right entanglement entropy does not diverge and it is in fact the 2d  thermodynamic entropy .

In this work another way to bipartite  the closed superstring Hilbert space will be explored. In D dimensions of spacetime the  light cone string Hilbert space is written as ${\cal H}= {\cal H}^1\otimes {\cal H}^2...{\cal H}^d$, where $d=D-2$. Clearly, for free strings propagating on a flat background, there is no entanglement between the different spacetime coordinates. However, it will be demonstrated herein that when the string is coupled to a thermal bath, i.e., strings in an ensemble of string states, the so-called extended density matrix can be defined and the extended entanglement between the different spacetime coordinates emerges. When entanglement is studied at finite temperature, it is challenging to differentiate between thermal and quantum fluctuations. A method based on Thermo Field Dynamics (TFD) was developed in reference \cite{Hashizume_2013} to overcome this challenge.  A salient feature of TFD is its intrinsic relationship with entanglement, which is a central focus of this study. The main go of the present work is to explore this relationship and calculate the extended entanglement entropy (EEE) for the type II Green-Schwarz superstring. 

The TFD is a canonical real time formalism. In general, in finite temperature real time formalism the degrees of freedom are duplicated. Unlike the Schwinger-Keldysh  formalism, where duplication is a consequence of the contour used in the path integral, in the TFD  the duplication is defined from the beginning in the context of the Tomita-Takesaki modular theory \cite{Takesaki:1970aki},  and the doublet fields are related to the  Gelfand-Naimark-Segal (GNS) representation induced by Kubo-Martin-Schwinger (KMS)  states \cite{lands} , \cite {Haag:1967sg}.
In addition to the implementation of a more rigorous apparatus for the study of thermal phenomena, the relationship between TFD and the GNS construction demonstrates the applicability of this formalism to the study of entanglement in a more general context. This helps to understand phenomena ranging from black hole entropy to quantum dissipation \cite{Dias:2019ezx}, \cite{BottaCantcheff:2017kys}. The following brief exposition aims to elucidate this relationship.

The core of the GNS construction is the GNS triple associated with the pair $(\mathcal{C}, \omega)$, composed of a $C^{\star}$-algebra  $\mathcal{C}$ and the state(a linear funtional) $\omega$ over $\mathcal{C}$. The GNS triple, $(\mathcal{H}_{\omega}, \pi_\omega,
\Omega_{\omega})$, is composed of the Hilbert space $\mathcal{H}_{\omega}$ , the $\pi_{\omega}$ representation of the  $C^{\star}$-algebra  and the cyclical vector $\Omega_{\omega}\in \mathcal{H}_\omega$. An important issue is the definition of state purity. The collection of all states in a $C^{\star}$-algebra $\mathcal{C}$ is a convex set. In fact, if $\omega_1$ and $\omega_2$ are states in $\mathcal{C}$, then, for all $\lambda \in [0, 1]$,
$\omega=\lambda\omega_1 + (1 - \lambda) \omega_2$ is also a state in $\mathcal{C}$. The state $\omega$ is said to be a mixed state if there are $\lambda \in  (0, 1)$ and $\omega_1, \omega_2$, with $\omega_1 \neq \omega$ and $\omega_2\neq\omega$, such that $\omega=\lambda\omega_1 + (1 - \lambda) \omega_2$. A state $\omega$ is said to be a pure state if it is not a mixed state. For example,  the KMS state $\omega_\beta$ is a mixed state and can be represented by $\omega_\beta(A)= Tr(\rho_\beta A) $, where A is a bounded linear operator and $\rho_\beta$ is the thermal density matrix at equilibrium temperature $T=\frac{1}{\beta}$. 
Another important ingredient of the algebraic approach is the Tomita-Takesaki modular theory, which establishes a one to one correspondence between an algebra $\mathcal{C}$  and its commutant $\mathcal{C}'$, and gives rise to the doubled degrees of freedom, in that $\mathcal{C}'$ is a copy of $\mathcal{C}$. The  modular theory of Tomita and Takesaki  play an important role to study KMS states  and establishes a formal structure to investigate the KMS condition in the thermodynamic limit. This was first pointed out by Haag, Hugenholtz and Winnink in Ref. \cite{Haag:1967sg}.


The relationship  between TFD and  the GNS construction  was laid down for the first time  in \cite{ojima81}  and it was further developed in  \cite {Dias:2020srq} to calculate the EE entropy for 2d  conformal theories. The following dictionary between algebraic terms and the language of TFD is established: the operator algebra will consist of the operator fields $\phi(x)$ and  the representative $\pi_{\omega}$ will be just $\phi(x)$. The cyclic vector $\Omega_\omega$ is the ket denoted by $\left|0(\beta)\right\rangle$, called the thermal vacuum.  The modular conjugate of $\phi(x)$ is called the tilde field, denoted by $\tilde{\phi}(x)$, so that the action of the Tomita operator J is: 
\begin{equation}
J\phi(x)J=\tilde{\phi}(x) \quad , \quad [\phi(x),\tilde{\phi}(x)]=0 \ .
\end{equation}
For for any operator $A(\phi)$ and $c\in \mathbb{C}$, the so-called tilde conjugation rules are derived by the action of the Tomita operator :
\begin{eqnarray}
(A_{i}(\phi)A_{j}(\phi))\widetilde{}&=&\widetilde{A}_{i}(\phi)\widetilde{A}_{j}(\phi) \ , \nonumber \\
(cA_{i}(\phi)+A_{j}(\phi))\widetilde{}&=&c^{\ast}\widetilde{A}_{i}(\phi)+\widetilde{A}_{j}(\phi) \ , \nonumber \\
(A_{i}^{\dagger }(\phi))\widetilde{}&=&(\widetilde{A}_{i})^{\dagger }(\phi) , \nonumber \\
 \widetilde {(\widetilde{A}_{i})(\phi)} &=&A_{i}(\phi) \ .\label{til}
\end{eqnarray}
The equilibrium TFD formalism is based on the GNS construction for a particular state: the KMS state $\omega_\beta$. In this case the statistical average of an operator $A(\phi)$ is written as the expected value of $A(\phi)$ in the thermal vacuum: 
\begin{equation}
\left\langle A(\phi)\right\rangle_{\beta}=\left\langle 0(\beta )\left|A(\phi)\right|0(\beta )\right\rangle \ .
\end{equation}
From the  Tomita-Takesaki theory: $J\Omega=\Omega$. This implies that the vacuum is tilde invariant. Time translations in the total Hilbert space ${\cal H}_{T}={\cal H}\otimes\widetilde{{\cal H}}$ is given by $\hat{H}=H-\tilde{H}$, and  the thermal vacuum is annihilated by $\hat{H}$. Finally,  by using the polar representation of the Tomita-Takesaki operator, the so-called thermal state condition is derived: 
\begin{equation}
e^{\beta\hat { H}/2}\tilde{\phi}(x)\left|0(\beta)\right\rangle=\phi^{\dagger}(x)\left|0(\beta)\right\rangle \ . \label{tscg}
\end{equation}

Note that all the properties presented so far do not depend on any particular characteristic of the field-theoretic system.  
Suppose that the hamiltonian  has a discrete spectrum and that it is possible to find a base $\left|\psi_n\right\rangle$ where the hamiltonian is diagonal. In this basis,  the thermal vacuum $\left|0(\beta)\right\rangle$ can be represented as
\begin{eqnarray}
\left|0(\beta)\right\rangle&=& \frac{1}{Z^{1/2}}\sum_{n}e^{-\frac{\beta E_n}{2}}|\psi_n,\tilde{\psi_n}\rangle, \label{tv}
 \end{eqnarray}
where Z is the normalization factor. This state can be seen as an entanglement state between the original and the tilde system, which is not surprising since the KMS state is a mixed state. 
In the context of ADS/CFT, the system and the tilde system can be interpreted as the left and right holographic CFTs of an extended AdS Schwarzschild geometry\cite{Maldacena:2001kr}. In this case, the thermal vacuum (\ref{tv}) is the state dual to the black hole. To conclude this subject, it is imperative to underscore the crucial interconnection between the irreducibility of the GNS representation and the purity of the state. The GNS representation $(\mathcal{H}_{\omega}, \pi_\omega,
\Omega_{\omega})$ is irreducible if and only if $\omega$ is a pure state on $\mathcal{C}$, and hence it cannot be written as a mixture of other states. Conversely, any mixed state on $\mathcal{C}$ induces a reducible GNS representation of the algebra. Finally, if the representation is found to be reducible, and thus if the state is pure, the commutant of the representation is trivial (i.e.  consists of scalar multiples of the identity). Thus, from the viewpoint of TFD, the existence of the tilde
system is closely related to the fact that the KMS state is not pure. In the ADS/CFT scenario this result allows us to understand the Kruskal extension of black holes from a quantum mechanical algebraic perspective. Also, as shown in \cite {Dias:2020srq}, the TFD technique can be used to study more general mixed states. In fact, the states in the tilde space in TFD play a role of tracers of the original ones. 

Let us redirect our focus to string theory. In the context of string theory at finite temperature, a notable issue pertains to the presence of a temperature beyond which the partition function exhibits a divergence. This temperature is referred to as the Hagedorn temperature. Despite the considerable research activity in the field of finite-temperature string theory, a comprehensive grasp of string physics at the Hagedorn temperature remains elusive. Noteworthy advancements in comprehending the Hagedorn phase can be attributed to the imaginary time formalism.
Within this formalism, the thermal state is delineated by compactifying Euclidean time on a circle, known as the thermal circle, with a radius equal to the inverse temperature in natural units. For string theory applications, however, this formalism introduces complications. The first complication arises from the fact that string theory contains gravity. In theories with gravity, the radius of the thermal circle becomes a dynamic field, rendering the notion of thermal equilibrium non-trivial \cite{Gross:1982cv}. Moreover, for closed strings, one must consider the winding modes surrounding the thermal circle. Above the Hagedorn temperature these modes become tachyonic. These tachyonic excitations, as previously discussed in \cite{Barbon:2001di}, \cite{Brustein:2022uft}, \cite{Salomonson:1988ac} and \cite{Salomonson:1985eq} encode the Hagedorn divergence and a long/short string transition. However, it is crucial to realize that time compatification in quantum field theory is only a means of incorporating the KMS condition.
 In thermal field dynamics (TFD), for instance, the KMS condition emerges from the thermal state condition (eq. \ref{tscg}),  obviating the necessity for time compactification. 
 In TFD, the partition function is merely a normalization factor in the thermal state, which is normalized to any temperature. From this perspective, the Hagedorn temperature emerges when calculating statistical averages of observables. Actually, as in hadron physics, the Hagedorn behavior in string theory is due to the exponential growth of states as a function of energy. If the Hagedorn behavior works in string theory as it does in hadron physics, then the true degrees of freedom of the theory at high temperatures may differ from those of the perturbative string. It is noteworthy that, in the applications referenced in the initial paragraph, EE serves as a measure of the degrees of freedom inherent to quantum theory. Consequently, within the framework of string theory, the investigation of the Hagedorn behavior of EE emerges as a subject of particular interest.  This is the main motivation of the present work, which is structured as follows: in section \ref{TFD} the application of TFD in string theory is revisited, but with a different approach than in other works \cite{adsnos},\cite{Nedel:2004gy},\cite{Abdalla:2005qs}; in section \ref{EDM} an important object for the calculation of the EEE  is introduced, the so-called extended density matrix; in section \ref{ET} the EEE for the bosonic string is calculated in detail and its UV behavior is studied; the results of section \ref{ET} are generalized in section \ref{S}, where the EEE is calculated for the type II superstring. Finally, the conclusions are presented in section \ref{conclusion}.

\section{TFD formalism in light cone finite temperature string theory.} \label{TFD}
Let's begin with a review of the TFD formalism as applied to string theory. In reference \cite{Abdalla:2005qs} it was pointed out that the natural environment to apply TFD in string theory is within the context of string field theory. Concerning the application of TFD in the first quantized string, in \cite{Nedel:2004gy} it was shown how to take into account the level matching condition of the closed string in the statistical averages. This procedure implies certain difficulties in reconciling the tilde conjugation rules with unitarity, which can be overcome by redefining the action of the Tomita Takesaki operator, as done in \cite{Abdalla:2003xg}. Here we present a different way of defining the light-cone Green Shwarz superstring thermal state. In order to fix the notation, some basic elements of the Green-Schawarz formulation had to be fixed \footnote{For an introduction to the Green Shwarz  formulation of superstring see ref. \cite{Green:1987sp} .}. 

In the Green-Schwarz formulation, the closed
superstring is defined through the worldsheet fields $X^\mu(\tau,\sigma)$
,$S_A^\alpha(\tau,\sigma)$ with $\mu=0,\ldots,10$, $A=1,2$ and
$\alpha=1,\ldots,16$ being 10d space-time vectors and two Majorana-Weyl
spinors, respectively. The 2d 
diffeomorphism invariance is used to go to the conformal gauge for the
worldsheet metric 
$
g_{ij}=e^{\phi}\, \left ( \begin{matrix} -1&0\cr 0& 1\cr
\end{matrix}\right )
$
and the residual conformal invariance
is  used to set 
$X^+(\tau,\sigma)= p^+\, \tau$.
Furthermore, the coordinate $X^-$ is expressed in terms of the transverse
degrees of freedom through the Virasoro constraint. In the fermion sector
the $\kappa$ symmetry is employed to gauge away one-half of the fermionic
degrees of freedom via the condition $\Gamma^+\, S_A^\alpha=0$. Finally, the light cone action for type IIB superstring is 

\beq
S&=&\frac{1}{2\pi\alpha'}\int_{-\infty}^{\infty}d\tau \int_0^{2\pi\alpha'p^+}
d\sigma\, \left(\partial_i X^I\,\partial_i X^I + i\, S_1^a(\partial_\tau+\partial_\sigma)S_1^a
+i\, S_2^a(\partial_\tau-\partial_\sigma)S_2^a \right)\,\, ,
\label{action}
\eea

where $I=1...8$ and  $a=1...8$,  so $S_1^a,S_2^a$ are $SO(8)$ spinors  of the same chirality. For type IIA, $S_2^a$ is replaced by $S_2^{\dot{a}}$. By defining $k_n=n/\alpha´p^+$, the solutions of the equations of motion with periodic boundary conditions
are 
\begin{eqnarray}
X^I&=& X_0^I + P_0^I\tau +
\sum_{n\neq 0}\frac{i}{k_n}\,(\,
 a^I_n\, e^{-ik_n(\tau-\sigma)}
+ \bar a^I_n\, e^{-ik_n(\tau+\sigma)}\, ) \, , \nonumber \\
S_1^a&=&\frac{1}{\sqrt{2}}\sum_{n\in \mathbb{Z}}S_n^a e^{-ik_n(\tau-\sigma)}\,,\,\,
S_2^a=\frac{1}{\sqrt{2}}\sum_{n\in \mathbb{Z}}\bar{S}_n^a e^{-ik_n(\tau+\sigma)}\,,
\end{eqnarray}
with  reality conditions: $a_{-n}^I=(a_n^I)^{\dagger}$, $S_{-n}^a=(S_n^a )^{\dagger}$.  The canonical quantization imply
\beq
\left[a_n^I,a_m^{J\dagger} \right]&=& \delta^{IJ}\delta_{nm}\,,\,\left[\bar a_n^I,\bar a_m^{J\dagger} \right]=\delta^{IJ}\delta_{nm} \nonumber \\
\{S_n^a ,S_m^{b\dagger}\}& =&\delta^{ab}\delta_{mn},\:\: \{\bar{S}_n^a ,\bar{S}_m^{b\dagger}\} =\delta^{ab}\delta_{mn}, \nonumber \\
\eea

In order to explore the dependence of the UV behavior of the entanglement entropy on the spacetime dimension, in the next sections we will not fix $d=8$ for the transverse dimensions, but will use only d. Then the thermal vacuum for the superstring is written in terms of the d-oscillators and the center of mass momentum, which makes it clear what kind of calculation will be done. The system will be bipartite into two sets of degrees of freedom: one system with the string transverse coordinates going from 1 to a number Q and another system going from Q+1 to d. We will then calculate the entanglement entropy between these two systems.The procedure for the bosonic sector will be presented in detail and then the results will be generalized to include the fermionic sector.

\subsection{The Light cone closed string ensemble.}
In the light cone, the killing vector $\frac{\partial}{\partial x^0}$ is written as $\frac{\partial}{\partial x^0}= \frac{\partial}{\partial x^+}+ \frac{\partial}{\partial x^-} $, so  the bosonic light cone thermal density matrix is:

\begin{equation}
\rho_b(\beta) = \frac{e^{-\beta(P^{+}+ P_b^{-})}}{Z_b}, \label{rob}
\end{equation}
where the partition function $Z_b$ is the normalization factor $Z_b=Tre^{-\beta(P^{+}+ P_b^{-})}$, $\beta$ is the inverse of temperature and $P_b^-$ is written in terms of the zero mode transversal momentum and the left and right number operators:
\beq
P_b^{-}&=& \frac{1}{\alpha'p^+}\left[\frac{P_{0}^2}{2}+\sum_{n=1}^\infty n(N_n +\bar{N}_n  ) +h_0\right],\nonumber \\
 N_n&=& \sum_Ia^{I \dagger}_na_n^I\:,\:\:\: \bar{N}=\sum_I\bar{a}^{I\dagger}_n\bar{a}_n^I \:,\:h_0=-\frac{1}{12}.\nonumber \\
 \eea
From this point on, to ease notation, $p$ will be used instead of $p_0$ . Note that $p^+,p^I$ are the eigenvalues of the $P^+,P^I$ operators and the term $h_0$ is the ordering constant. Although the light cone gauge solves the two
worldsheet reparametrization constraints, there remains a single consistency condition to be imposed on the closed string Hilbert space, which is related to the circle isometry. So, the physical closed string states  $ |\Phi\rangle$ are annihilated  by the worldsheet translation generator
\begin{equation}
(N-\bar N)|\Phi\rangle=0 \label{constr}
\end{equation}
where $N= \sum\limits_{n=1}^\infty N_n$. This is known as the level matching condition.
By defining the projector 
\be
\int_{-\frac{1}{2}}^{\frac{1}{2}} d\lambda e^{2\pi i (N-\bar N)}\:, \label{proj}
\ee
 the normalization factor of the light cone density  matrix (the partition function) is written as
   \begin{equation}
Z_b=\int  dp^{+}d\lambda d^{d}pe^{-\beta p^+}e^{-\beta \frac{p^2}{2\alpha' p^+}} z_b(\beta,\lambda)\label{Z},
\end{equation}
where
\begin{eqnarray}
z_b(\beta,\lambda) = \sum_{\{n^I_i\},\{\bar{n}^I_i\}}\left\langle \{n^I_i\},\{\bar{n}^I_i\},p^+,p^I\right|e^{-\beta P_-+2\pi i\lambda(N-\bar{N})}\left|\{n^I_i\},\{\bar{n}^I_i\},p^+,p^I\right\rangle 
\end{eqnarray}
with $\{n^I_i\} =n^1_1,\dots,n^d_1,\dots n^1_\infty,\dots n^d_\infty$ and the same for $\{\bar{n}^I_i\}$. The sum $\sum\limits_{\{n_i\}}$ means 
\begin{equation} 
\sum_{\{n^I_i\}}=\sum_{n^1_1,\dots{n}^d_1}\dots\sum_{n^1_{\infty},{n}^d_\infty}.
\end{equation}
The projector (\ref{proj}) ensures that the trace is taken over the physical states of the closed string. Note that
$P^+|\{n^I_i\},p^+,p^I\rangle=p^+|\{n^I_i\},p^+,p^I\rangle$  and $P^2|\{n^I_i\},p^+,p^I\rangle=p^2|\{n^I_i\},p^+,p^I\rangle$. It is well known that $Z_b$ starts to diverge at the Hagedorn temperature 
\begin{equation}
T_H= \frac{1}{\beta_H}= \sqrt{\frac{3}{\pi d \alpha^\prime}}. \label{hage}
\end{equation}
Let's turn our attention to TFD. Given the density matrix, the statistical average of a worldsheet operator $O$ in an ensemble of bosonic string states in thermal equilibrium is
\begin{equation}
    \langle O\rangle_\beta=Tr\langle \rho_b(\beta) O\rangle \label{media}
\end{equation}
and the trace is taken in the same way as in (\ref{Z}). The main idea of the TFD is to write this statistical average as the expectation value of $O$ in a pure state, named  thermal vacuum. In general, the thermal vacuum for free theories is constructed through a Bogoliubov transformation which entangles the original system with its copy. In flat space string theory this process presents difficulties in relation to the continuous zero mode. Also, for applications in closed string, the level matching condition must be implemented in the statistical average. In ref \cite{Nedel:2004gy} the pp wave closed string thermal vacuum was defined from a usual thermal Bogoliubov transformation  and the constraint (\ref{constr}) was imposed by redefining the Hamiltonian as in \cite{Ojima:1981ma}.  Here, the level matching condition is taken into account in the definition of the expectation value. Let's start duplicating the degrees of freedom  by defining  the notation
\begin{equation}
    |\{n^I_i\},\{\bar{n}^I_i\},p^+,p^I\rangle\rangle=  |\{n^I_i\},\{\bar{n}^I_i\},p^+,p^I\rangle\otimes |\{\widetilde{n}^I_i\},
    \{\widetilde{\bar{n}}^I_i\},\widetilde{p}^+,\widetilde{p}^I\rangle.
\end{equation}
From the tilde conjugation rules, it can be shown that the tilde Hilbert space is in fact a Hilbert space of a string propagating backwards in time  \cite{Abdalla:2004dg}. Next, we define the following state with fixed momentum
\begin{equation}
\left|\Psi(\beta),p^+,p^I,\tilde{p}^+,\tilde{p}^I\right\rangle =\frac{e^{-\frac{\beta}{2}(\frac{ p^2+2h_0}{2\alpha'p^+}+ p^+)}}{\sqrt{Z_b}}\sum_{\{n^I_i\},\{\bar {n}^I_i\}} e^{-\frac{\beta}{2}\sum\limits_{I=1 }^d \sum\limits_{j = 1}^\infty j(n^I_j+\bar{n}^I_j)}|\{n^I_j\},\{\bar{n}^I_j\},p^+,p^{I}\rangle\rangle .\label{fixedp} 
\end{equation}
Using the creation operators for the original and the tilde Hilbert space, the state (\ref{fixedp}) can be written as
\be
\left|\Psi(\beta),p^+,p^I,\tilde{p}^+,\tilde{p}^I\right\rangle= \frac{e^{-\frac{\beta}{2}(\frac{ p^2+2h_0}{2\alpha'p^+}+ p^+)}}{\sqrt{Z_b}}e^{\sum\limits_{n = 1}^\infty G_n(\beta)}|0,\bar{0},p^+,p^I\rangle\rangle \label{termaosc}
\ee
where
\be 
G_n = e^{-\frac{\beta n}{2}}\sum\limits_{I=1}^{d}\left(a_n^{I\dagger}\tilde{a}_n^{I\dagger}+ \bar{a}_n^{I\dagger}\tilde{\bar{a}}_n^{I\dagger}\right).
\ee
The string thermal vacuum is just the integral over the momenta
\begin{equation}
\left|\Psi(\beta)\right\rangle =\int dp^+d^dp^I\left|\Psi(\beta),p^+,p^I,\tilde{p}^+,\tilde{p}^I\right\rangle,
\end{equation}
such that the statistical average defined in ({\ref{media}}) can be written as 

\begin{eqnarray}
\left\langle O\right\rangle_\beta = \int d\lambda \langle \Psi(\beta)|O e^{2\pi i\lambda( N-\bar{N})} |\Psi(\beta)\rangle,
\end{eqnarray}
where the level matching condition is introduced in the definition of the TFD vacuum expectation value.  As an example, let us calculate the expectation value of the light cone  Hamiltonian ($P_b^-$) in the thermal state. It is easy to show that
 \begin{equation}
  \int d\lambda \left\langle \Psi(\beta)\right|P_b^- e^{2\pi i\lambda( N-\bar{N})}\left|\Psi(\beta)\right\rangle =  -\frac{\partial}{\partial \beta}\ln Z_b ,
 \end{equation}
 which  is the thermal energy. Next, let's rewrite  the state (\ref{fixedp}) as
 \be
 \left|\Psi(\beta),p^+,p^I,\tilde{p}^+,\tilde{p}^I\right\rangle=e^{-\frac{K_\beta}{2}} I\:,\:\:\: I=\sum\limits_{\{n_i\}}|\{n^I_i\},\{\bar{n}^I_i\},p^+,p^I\rangle\rangle
 \ee
where 
 \be
 K_\beta=\beta P_b^-+\ln Z_b \label{soperator}
 \ee
 is an operator whose expectation value in $\left|\Psi(\beta)\right\rangle$  provides the thermodynamic entropy:
 \be
 S_\beta=\int d\lambda \langle \Psi(\beta)|K_\beta e^{2\pi i\lambda( N-\bar{N})} |\Psi(\beta)\rangle .
 \ee
  It is worth noting that the dependence of $ \left|\Psi(\beta),p^+,p^I,\tilde{p}^+,\tilde{p}^I\right\rangle$ on $\beta$ comes from the entropy operator; thus, $K_{\beta}$ generates the thermal state. In the next
section another way to calculate  entropy in the TFD formalism will be shown, where it will be clear that $S_\beta$ measures
the entanglement between the original and the tilde system.

\section{Extended Density Matrix.} \label{EDM}
An important object in TFD  is the so-called extended density matrix. In \cite{Hashizume_2013} it was shown that the extended density matrix allows us to distinguish intrinsic quantum
entanglement from the thermal fluctuations included in the definition of the quantum entanglement at finite temperatures. For the closed string  the extended density matrix is


\begin{eqnarray}
\rho^e &=&\left|\Psi(\beta)\right\rangle\left\langle \Psi(\beta)\right|\nonumber \\
&=&
\frac{1}{Z_b}\int d^dpd^dqdp^+dq^+ e^{ - \frac{\beta  }{2}\, \left(\sum\limits_{j,I}  j\,(n^I_j+\bar{n}^I_j+ {m^I_j}+\bar{m}^I_j) \right)}e^{-\frac{\beta}{2}(p^++q^+)}e^{-\frac{\beta}{4\alpha'}(\frac{p^2}{p^+}+\frac{q^2}{q^+}+2h_0)} \,{| \{ {n_j^I}\},p^+,p^I  \rangle\rangle\langle\langle \{ {m_j^I}\},q^+,q^I|}. \nonumber \\
\end{eqnarray}

If we trace the tilde variables as follows:
\begin{equation}
 \tilde {T_r} \rho^e=\int d\widetilde{l}^+\int d^d \widetilde{l}^I \sum\limits_{\{\tilde{n}^I_i\},\{\tilde{\bar{n}}^I_i\}}  \langle \{ \widetilde{n}^I _i\} , \{ \widetilde {\bar{n}}^I _i\},\widetilde{l}^+,\widetilde{l}^I|\rho^e| \{ \widetilde{n}^I _i\} ,\{\widetilde {\bar{n}}^I _i\},\widetilde{l}^+,\widetilde{l}^I\rangle  ,
 \end{equation}
 we get
 \begin{eqnarray}
 \tilde {T_r} \rho^e &=&\frac{1}{Z_b}\int d^dpdp^+\sum\limits_{\{ n^I_i\},\{\bar{n}^I_i\}}  e^{-\beta\sum\limits_{i,I}(n_i^I+\bar{n}_i^I)}
e^{-\beta(\frac{p^2+2h_0}{2\alpha^\prime p^+}+p^+)}|\{ {n_i^I}\},\{\bar{n}_i^I\},p^+,p^I\rangle\langle \{n_i^I\},\{\bar{n}_i^I\},p^+,p^I| \nonumber \\
&=&\rho_b(\beta),
 \end{eqnarray}
 which is the thermal equilibrium  density matrix (\ref{ro}) in number basis. The calculation of the von Neumann entropy associated with this density matrix yields the closed string thermodynamic entropy:
\begin{equation}
S_{\beta}= Tr\rho_b(\beta)\ln\rho_b(\beta),
 \end{equation}
where in the trace it is necessary to introduce the level matching condition as in (\ref{Z}). Notice that $S_{\beta}$ is a measure of the entanglement between the original and the tilde system. So, based on \cite{Abdalla:2004dg}, $S_{\beta}$  measures the entanglement between states of the original string with another that propagates backwards in time.

\section{The Extended Entanglement Entropy} \label{ET}
Let's now define a different trace. We are going to trace over a subset of oscillators and its correspondent tilde defined in different space time coordinates. 
To do this,  the  system is birpatited as follows
\begin{equation}
\left\{ {{n_i^I}} \right\}_{I= 1}^d = \left\{ {{n_i^B}} \right\}_{B= 1}^Q\bigcup {\left\{ {{n_i^A}} \right\}_{A = Q + 1}^d} \,,\quad Q \le d - 1. \label{parti}
\end{equation}
 The reduced density matrix is defined by calculating the trace over the B system 

\begin{eqnarray}
\rho_A&=&  Tr_B\rho^e=\sum_{\{l^B_i\},\{\bar{l}^B_i\}}\langle\langle \{l^B_i\},\{\bar{l}^B_i\}| \rho^e(p^+,q^+,q^I,p^I)|\{l^B_i,\{\bar{l}^B_i\}\}\rangle\rangle \nonumber \\
&=&\frac{1}{Z_b}e^{-\frac{\beta}{2}(p^++q^+)}e^{-\frac{\beta}{4\alpha'}(\frac{p^2}{p^+}+\frac{q^2}{q^+}+2h_0)}\prod\limits_{n=1}^\infty\left[\frac{1}{1-e^{-\beta\frac {n}{\alpha'p^+}}}\right]^{2Q} \nonumber\\
&\times &\sum\limits_{\{n_i^A\}, \{\bar{n}_i^A\},\{m_i^A\}, \{\bar{m}_i^A\}}e^{-\frac{\beta}{2}\sum\limits_{A=Q+1}^d\sum\limits_{i=1}^\infty i(n_i^A+m_i^A+\bar{n}_i^A+\bar{m}_i^A)}
|\{n_i^A\},\{\bar{n}_i^A\},q^+,q^I,\rangle\langle \{m_i^A\},\{\bar{m}_i^A\},p^+,p^I|,\nonumber \\
\end{eqnarray}
and the entanglement entropy is
\begin{equation}
S=-Tr\rho_A\ln\rho_A .
\end{equation}
 By imposing the level matching condition on the trace, the entanglement entropy becomes 
\begin{eqnarray}
S=-\int d\lambda\int d^{d}ldl^+\sum_{\{n_i^A\},\{\bar{n}_i^A\}}\langle  \{n_i^A\},\{\bar{n}_i^A\},p,p^+|\rho_A\ln\rho_A e^{2\pi i(N-\bar{N})}|\{n_i^A\},\{\bar{n}_i^A\},p,p^+\rangle .
\end{eqnarray}
In order to write the entanglement entropy in terms of modular functions, let's define 

\begin{eqnarray}
\tau &=&\tau_1+i\tau_2 ,\nonumber \\
\tau_1&=& \lambda, \:\:\: \tau_2=\frac{\beta}{2\pi\alpha'p^+} ,\nonumber \\
\end{eqnarray}
which in general is related to the torus moduli space parameter. Now, the UV limit corresponds to $\tau_2\rightarrow 0$. Defining $q=e^{2\pi i\tau}$,$q^\prime = q(\tau_1=0)$ , the entropy is written as
\begin{equation}
S=\frac{\beta}{2\pi\alpha'Z_b}\frac{\int d\tau_1 d\tau_2}{\tau_2^2}\tau_2^{-\frac{d}{2}}F(\tau_1,\tau_2)G(\tau_1,\tau_2)\left[d+ 
\ln\frac{e^{-\frac{\pi(d-Q)\tau_2}{6}}F(\tau_1,\tau_2)}{Z}+\frac{\beta}{4}(\frac{\Psi_q(1)}{\ln q} + \frac{\Psi_{\bar{q}}(1)}{\ln \bar{q}} ) \right ]  ,
\end{equation} 
where $\Psi_q(1 )$ is the q-Pollygamma function 
\be
\Psi_q(1)=  \sum n\frac{q^n}{1-q^n} 
\ee
  and the functions $F(\tau_1,\tau_2)$ and $G(\tau_1,\tau_2)$  are written in terms of the Eta function 
\begin{eqnarray}
F(\tau_1,\tau_2)&=& e^{-\frac{\beta^2}{2\pi\tau_2}}e^{-\frac{\pi Q\tau_2}{6}}\left(\prod\frac{1}{1-e^{-2\pi n\tau_2}}\right)^{2Q}= |\eta(q^\prime)|^{-2Q}, \nonumber \\
G(\tau_1,\tau_2)&=& {e^{-\frac{\pi(d-Q)\tau_2}{6}}\left|\prod\frac{1}{1-(q)^n}\right|^{2(D-Q)}}= |\eta(q)|^{-2(d-Q)} .\nonumber \\
\end{eqnarray}

The study of UV behavior is typically facilitated by using the modular properties of the eta function:
\begin{equation}
    \eta(-\frac{1}{\tau})= \sqrt{-i\tau}\eta(\tau) .\label{etapr}
\end{equation}
 The term involving $ \ln{e^{-\frac{\pi(d-Q)\tau_2}{6}}F(\tau_1,\tau_2)}$ is irrelevant in the UV limit. Let's  focus on the term
\begin{equation}
       \frac{{\int d\tau_2}\tau_2^{-\frac{d}{2}-2}F(\tau_1,\tau_2)G(\tau_1,\tau_2)}{Z_b} . \label{comb}
\end{equation}
In order to study the Hagedorn behavior we fix $\tau_1=0$ and use (\ref{etapr}). In the UV limit the dominant term of the numerator of (\ref{comb}) behaves as
\begin{equation}
\int d\tau_2\tau_2^{\frac{d}{2}-2}exp \left[\frac{d\pi}{3\tau_2}-\frac{\beta^2}{2\pi\alpha^\prime \tau_2}\right], \label{dominant}
\end{equation}
and it begins to diverge exactly at the Hagedorn temperature $T_H$ defined at (\ref{hage}). This is precisely the same exponential behavior as the denominator $Z_b$. Thus the term (\ref{comb}) does not exhibit Hagedorn behavior.
Let's now analyze the terms involving the functions $\Psi_q(1)$ and $\Psi_{\bar{q}}(1)$. Note that this term contributes less as the temperature increases; so, this is a typical entanglement term. It has the following UV behavior \cite{doi:10.1142/S1793042117501135}
\begin{equation}
\lim_{\tau_ 2 \rightarrow 0}\frac{\Psi(q)}{\ln q} \approx \frac{1}{\tau_2^2}. \label{pureen}
\end{equation}
Then, combining this term with the powers of $\tau$ that come from equation (\ref {dominant}), we conclude that the terms proportional to $\beta^2$ in the EE are UV finite at the critical dimension $d = 8$; also, this behavior is independent of temperature.

Finally, we conclude that the only term that exhibits Hagedorn behavior is the term $\ln Z_b$. Thus the entanglement entropy begins to diverge at the Hagedorn temperature $T_H$.  Note that, owing to the term $\tau^{\frac{d}{2}-2}$, for the temperatures below the Hagedorn temperature the entanglement entropy is finite in the UV limit for $d \geq 8$.

\section{ The Extended Entanglement Entropy for Superstring. }\label{S}

The generalization of the steps developed in the previous section to include fermionic degrees of freedom is straightforward. For the type II B superstring, after fixing the light cone gauge and kappa symmetry, the  thermal density matrix is:

\begin{equation}
\rho(\beta) = \frac{e^{-\beta(P^{+}+ P^{-})}}{Z}, \label{ro}
\end{equation}
where
\be
P^{-}= \frac{1}{\alpha'p^+}\left[\frac{P^2}{2}+\sum_{n=1}^\infty n(N_n +\bar{N}_n +{\cal N}_n+\bar{{\cal N}}_n )\right]
\ee
and ${\cal N}_n$, $\bar{\cal {N}}_n$ are the left and right moving fermionic numbers operator:
\be
{\cal N}_n =\sum_{a=0}^d S_n^{a\dagger}S_n^a \:,\:\:\:\bar{\cal {N}}_n =\sum_{a=0}^d\bar{S}_n^a{}^\dagger \bar{S}_n^a .
\ee
The trace of the density matrix is taken in the same way as before  and the level matching condition is imposed by replacing projector  (\ref{proj} ) with 
\be
\int_{-\frac{1}{2}}^{\frac{1}{2}} d\lambda e^{2\pi i (N+{\cal N}-\bar N -\bar{{\cal N}})}\:, \label{proj2}
\ee
where ${\cal N}= \sum\limits_{n=0}^\infty {\cal N}_n$.  The trace of the fermionic oscillator part of the density matrix is
\be
z_f(\tau_1,\tau_2)=Tre^{-\frac{\beta}{\alpha^{\prime}p^+}( {\cal N}+\bar{{\cal N}} )}e^{2\pi i\lambda(\cal{N}-\bar{{\cal N}})}= \left(|q|^{\frac{1}{12}}\prod_{n=1}^\infty  \bigg| 1+q^n\bigg|^2 \right)^{d}
\ee
and the total partition function for the superstring can be written as
\begin{eqnarray}
Z(\beta)&=& \int  dp^{+}d\lambda d^{d}pe^{-\beta p^+}e^{-\beta \frac{p^2}{2\alpha' p^+}}z_f(\tau_1,\tau_2)z_b(\tau_1,\tau_2) \nonumber \\
&=& = \frac{\beta}{2\pi l_s} \int_0^\infty \frac{d\tau_2}{\tau_2{}^2}  
\int d\tau_1  
\exp(-\frac{\beta^2}{2\pi\alpha'\tau_2}) 2^{-16}\tau_2^{-\frac{d}{2}}\left|\Theta_2(0,\tau)   
\eta(\tau)^{-3}\right|^{2d}\, .
\end{eqnarray}
where the $\Theta_2(0,\tau) $ function is
\be
\Theta_2(0,\tau) = 2q^{\frac{1}{8}}\prod_{n=1}^{\infty}(1+q^n )^2(1-q^n).
\ee
Now the Hagedorn temperature is 
\be
T_H=4/(d\pi \sqrt{\alpha^\prime} ).  
\ee
The thermal vacuum  of the closed superstring is defined just by replacing $G_n$ in (\ref{termaosc}) with
\be 
{\cal G}_n= e^{-\frac{\beta n}{2}}\left[\sum\limits_{I }^d\left(a_n^{I\dagger}\tilde{a}_n^{I\dagger}+ \bar{a}_n^{I\dagger}\bar{\tilde{a}}_n^{I\dagger}\right)+ \sum_{a=0}^{d}\left(S_n^{a\dagger}\tilde{S}_n^{a\dagger}+ \bar{S}_n^{a\dagger}\tilde{\bar S}_n^{a\dagger}\right)\right].
\ee
Following the same procedure described in the previous sections, the entanglement entropy for the closed superstring is
\begin{equation}
S=\frac{\beta}{2\pi\alpha'Z}\frac{\int d\tau_1 d\tau_2}{\tau_2^2}\tau_2^{-\frac{d}{2}}{\cal F}(q){\cal G}(q)\left[d+ 
\ln\frac{{\cal F}(q)}{Z}+\frac{\beta}{4}(\frac{\Psi_{q^2}(\frac{1}{2})}{\ln q} + \frac{\Psi_{\bar{q}^2}(\frac{1}{2})}{\ln \bar{q}} ) \right ] ,  
\end{equation}
where
\be
{\cal F}(q)=\left|\Theta_2(0,q)   
\eta(q)^{-3}\right|^{2Q},\:\:\: {\cal G}(q)=\left|\Theta_2(0,q^{\prime})   
\eta(q^{\prime})^{-3}\right|^{2(d-Q)}
\ee
and it was used 
\be
\sum\frac{nq^n}{1+q^n} = -\frac{\Psi_q(1)}{2\ln q} + \frac{\Psi_{q^2}(\frac{1}{2})}{4\ln q}.
\ee
Again, by using the properties of the q functions near $q=1$, we get

\begin{equation}
\lim_{\tau\rightarrow 0}\frac{{\Psi}_{q^2}(\frac{1}{2})}{\ln q}\approx \frac{1}{\tau ^2}
\end{equation}
and this term has the same UV behavior as the bosonic string.  The following discussion will concentrate on the term
\begin{eqnarray}
\int d\tau_2\frac{\tau_2^{-\frac{d}{2}-2}{\cal F}(q){\cal G}(q)}{Z} \:. \label{superH} 
\end{eqnarray}

Using $\Theta_4(q^2)= \sqrt{\Theta_3(q)\Theta_4(q)}$ and $2\eta^3(q)=\Theta_2(0,q)\Theta_3(0,q)\Theta_4(0,q)$, we have 
\begin{equation}
\lim_{\tau\rightarrow 0}\left[\eta(q)^{-3}\right|^{2Q}\left|\Theta_2(0,q^{\prime})   
\eta(q^{\prime})^{-3}\right|^{2(d-Q)}= \lim_{\tau\rightarrow 0}|\Theta_4(0,q^2)|^{2d}.
\end{equation}
   Next, invoking the modular properties of $\Theta_4$  
\be  
\Theta_4(0,\tau)=(-i\tau)^{-\frac 12}\Theta_2(0,-\frac 1\tau) , 
\ee  
 the numerator of equation (\ref{superH}) has the following UV behavior:  
\be  
\tau_2^{\frac{d}{2}-2} e^{-\frac{\beta^2}{2\pi\alpha'\tau_2}} e^{\frac{d\pi}{4\tau_2}}.  
\ee  

Therefore, analogous to the bosonic case, the numerator has the same exponential behavior as the denominator and begins to diverge at the same temperature.  The only temperature dependent divergence of the superstring entanglement entropy comes from the term $\ln Z$. Also, for temperatures below the Hagedorn temperature, the entanglement entropy for the superstring is finite in the UV limit. 

\section{Conclusions.} \label{conclusion}
A natural bipartition of the  string Hilbert space is achieved through the space coordinates. Evidently, for a free string at zero temperature there is no entanglement between the coordinates. However, it has been demonstrated that at finite temperature it is possible to define the so called extended density matrix, which exhibits entanglement  between the oscillators defined in different coordinates. The extended entanglement entropy for the closed superstring was determined through the TFD approach developed in \cite{Hashizume_2013} and expressed in terms of modular functions.
 It is noteworthy that this approach had previously been employed in  \cite{Dimov:2017ryz} for the bosonic string propagating in a pp wave background. However, a substantial approximation was employed, wherein only the zero mode was analyzed. Furthermore, in the case of pp waves, the zero mode does not possess a continuous spectrum.

It has been shown that the entropy can be expressed as the sum of two contributions, namely the $\beta$ and $\beta^2$ order, and the modular properties of the entropy have been used to study its ultraviolet (UV) behavior. The sector proportional to $\beta^2$ is shown to be finite in the UV regime for any temperature. This sector corresponds to typical quantum fluctuations and decouples at high temperatures. The sector proportional to $\beta$ diverges at the Hagedorn temperature, but is UV finite below the Hagedorn temperature. Therefore, it can be concluded that the Hagedorn behavior is exclusively due to thermal fluctuations and is not affected by quantum entanglement. Indeed, one of the advantages of the method used here is that it allows one to separate typical thermal fluctuations from quantum fluctuations due to entanglement. The finite UV behavior of entanglement entropy differs from the conventional behavior of entanglement entropy in particle field theories. This can be attributed to the fact  that the string is an extended object and the parameter $\alpha^{\prime}$ functions as a natural cutoff.  However, for the type of EE calculated here, at temperatures
below the Hagedorn temperature,  the entanglement entropy is finite in the UV limit only for $d \geq 8$. This UV dependence on dimension is a key motivation for extending the calculations performed here to scenarios where some dimensions are compactified. This extension will be pursued in future work. An interesting extension of this work is to investigate under what circumstances this type of entanglement can occur for strings at zero temperature. A promising avenue of exploration in this regard lies in the study of closed strings in plane wave backgrounds with a homogeneous NS-NS three-form field strength \cite{Blau:2003rt}. This background is particularly advantageous due to the solvable nature of the sigma model and the existence of coupling between different coordinates, which facilitates a comprehensive study of potential coordinate entanglement scenarios for the string propagating in this geometry.

\bibliography{biblio}

\begin{thebibliography}{41}
\expandafter\ifx\csname natexlab\endcsname\relax\def\natexlab#1{#1}\fi
\expandafter\ifx\csname bibnamefont\endcsname\relax
  \def\bibnamefont#1{#1}\fi
\expandafter\ifx\csname bibfnamefont\endcsname\relax
  \def\bibfnamefont#1{#1}\fi
\expandafter\ifx\csname citenamefont\endcsname\relax
  \def\citenamefont#1{#1}\fi
\expandafter\ifx\csname url\endcsname\relax
  \def\url#1{\texttt{#1}}\fi
\expandafter\ifx\csname urlprefix\endcsname\relax\def\urlprefix{URL }\fi
\providecommand{\bibinfo}[2]{#2}
\providecommand{\eprint}[2][]{\url{#2}}

\bibitem[{\citenamefont{Fendley et~al.}(2007)\citenamefont{Fendley, Fisher, and Nayak}}]{Fendley:2006gr}
\bibinfo{author}{\bibfnamefont{P.}~\bibnamefont{Fendley}}, \bibinfo{author}{\bibfnamefont{M.~P.~A.} \bibnamefont{Fisher}}, \bibnamefont{and} \bibinfo{author}{\bibfnamefont{C.}~\bibnamefont{Nayak}}, \bibinfo{journal}{J. Statist. Phys.} \textbf{\bibinfo{volume}{126}}, \bibinfo{pages}{1111} (\bibinfo{year}{2007}), \eprint{cond-mat/0609072}.

\bibitem[{\citenamefont{Levin and Wen}(2006{\natexlab{a}})}]{PhysRevLett.96.110405}
\bibinfo{author}{\bibfnamefont{M.}~\bibnamefont{Levin}} \bibnamefont{and} \bibinfo{author}{\bibfnamefont{X.-G.} \bibnamefont{Wen}}, \bibinfo{journal}{Phys. Rev. Lett.} \textbf{\bibinfo{volume}{96}}, \bibinfo{pages}{110405} (\bibinfo{year}{2006}{\natexlab{a}}), \urlprefix\url{https://link.aps.org/doi/10.1103/PhysRevLett.96.110405}.

\bibitem[{\citenamefont{Levin and Wen}(2006{\natexlab{b}})}]{Levin:2006zz}
\bibinfo{author}{\bibfnamefont{M.}~\bibnamefont{Levin}} \bibnamefont{and} \bibinfo{author}{\bibfnamefont{X.-G.} \bibnamefont{Wen}}, \bibinfo{journal}{Phys. Rev. Lett.} \textbf{\bibinfo{volume}{96}}, \bibinfo{pages}{110405} (\bibinfo{year}{2006}{\natexlab{b}}), \eprint{cond-mat/0510613}.

\bibitem[{\citenamefont{Holzhey et~al.}(1994)\citenamefont{Holzhey, Larsen, and Wilczek}}]{Holzhey:1994we}
\bibinfo{author}{\bibfnamefont{C.}~\bibnamefont{Holzhey}}, \bibinfo{author}{\bibfnamefont{F.}~\bibnamefont{Larsen}}, \bibnamefont{and} \bibinfo{author}{\bibfnamefont{F.}~\bibnamefont{Wilczek}}, \bibinfo{journal}{Nucl. Phys. B} \textbf{\bibinfo{volume}{424}}, \bibinfo{pages}{443} (\bibinfo{year}{1994}), \eprint{hep-th/9403108}.

\bibitem[{\citenamefont{Calabrese and Cardy}(2004)}]{Calabrese:2004eu}
\bibinfo{author}{\bibfnamefont{P.}~\bibnamefont{Calabrese}} \bibnamefont{and} \bibinfo{author}{\bibfnamefont{J.~L.} \bibnamefont{Cardy}}, \bibinfo{journal}{J. Stat. Mech.} \textbf{\bibinfo{volume}{0406}}, \bibinfo{pages}{P06002} (\bibinfo{year}{2004}), \eprint{hep-th/0405152}.

\bibitem[{\citenamefont{Ryu and Takayanagi}(2006)}]{RT}
\bibinfo{author}{\bibfnamefont{S.}~\bibnamefont{Ryu}} \bibnamefont{and} \bibinfo{author}{\bibfnamefont{T.}~\bibnamefont{Takayanagi}}, \bibinfo{journal}{Phys. Rev. Lett.} \textbf{\bibinfo{volume}{96}}, \bibinfo{pages}{181602} (\bibinfo{year}{2006}), \eprint{hep-th/0603001}.

\bibitem[{\citenamefont{Rangamani and Takayanagi}(2017)}]{Rangamani_2017}
\bibinfo{author}{\bibfnamefont{M.}~\bibnamefont{Rangamani}} \bibnamefont{and} \bibinfo{author}{\bibfnamefont{T.}~\bibnamefont{Takayanagi}}, \emph{\bibinfo{title}{Holographic Entanglement Entropy}} (\bibinfo{publisher}{Springer International Publishing}, \bibinfo{year}{2017}), ISBN \bibinfo{isbn}{9783319525730}, \urlprefix\url{http://dx.doi.org/10.1007/978-3-319-52573-0}.

\bibitem[{\citenamefont{Dias et~al.}(2021{\natexlab{a}})\citenamefont{Dias, Nedel, and Senise}}]{Dias:2020srq}
\bibinfo{author}{\bibfnamefont{M.}~\bibnamefont{Dias}}, \bibinfo{author}{\bibfnamefont{D.~L.} \bibnamefont{Nedel}}, \bibnamefont{and} \bibinfo{author}{\bibfnamefont{C.~R.} \bibnamefont{Senise}}, \bibinfo{journal}{Int. J. Mod. Phys. A} \textbf{\bibinfo{volume}{36}}, \bibinfo{pages}{2150092} (\bibinfo{year}{2021}{\natexlab{a}}), \eprint{2007.05365}.

\bibitem[{\citenamefont{He et~al.}(2015)\citenamefont{He, Numasawa, Takayanagi, and Watanabe}}]{He:2014gva}
\bibinfo{author}{\bibfnamefont{S.}~\bibnamefont{He}}, \bibinfo{author}{\bibfnamefont{T.}~\bibnamefont{Numasawa}}, \bibinfo{author}{\bibfnamefont{T.}~\bibnamefont{Takayanagi}}, \bibnamefont{and} \bibinfo{author}{\bibfnamefont{K.}~\bibnamefont{Watanabe}}, \bibinfo{journal}{JHEP} \textbf{\bibinfo{volume}{05}}, \bibinfo{pages}{106} (\bibinfo{year}{2015}), \eprint{1412.5606}.

\bibitem[{\citenamefont{Mertens et~al.}(2016)\citenamefont{Mertens, Verschelde, and Zakharov}}]{Mertens:2015adr}
\bibinfo{author}{\bibfnamefont{T.~G.} \bibnamefont{Mertens}}, \bibinfo{author}{\bibfnamefont{H.}~\bibnamefont{Verschelde}}, \bibnamefont{and} \bibinfo{author}{\bibfnamefont{V.~I.} \bibnamefont{Zakharov}}, \bibinfo{journal}{Phys. Rev. D} \textbf{\bibinfo{volume}{93}}, \bibinfo{pages}{104028} (\bibinfo{year}{2016}), \eprint{1511.00560}.

\bibitem[{\citenamefont{Balasubramanian and Parrikar}(2018)}]{Balasubramanian:2018axm}
\bibinfo{author}{\bibfnamefont{V.}~\bibnamefont{Balasubramanian}} \bibnamefont{and} \bibinfo{author}{\bibfnamefont{O.}~\bibnamefont{Parrikar}}, \bibinfo{journal}{Phys. Rev. D} \textbf{\bibinfo{volume}{97}}, \bibinfo{pages}{066025} (\bibinfo{year}{2018}), \eprint{1801.03517}.

\bibitem[{\citenamefont{Balasubramanian et~al.}(2012)\citenamefont{Balasubramanian, McDermott, and Van~Raamsdonk}}]{Balasu}
\bibinfo{author}{\bibfnamefont{V.}~\bibnamefont{Balasubramanian}}, \bibinfo{author}{\bibfnamefont{M.~B.} \bibnamefont{McDermott}}, \bibnamefont{and} \bibinfo{author}{\bibfnamefont{M.}~\bibnamefont{Van~Raamsdonk}}, \bibinfo{journal}{Phys. Rev. D} \textbf{\bibinfo{volume}{86}}, \bibinfo{pages}{045014} (\bibinfo{year}{2012}), \eprint{1108.3568}.

\bibitem[{\citenamefont{Pando~Zayas and Quiroz}(2016)}]{z2}
\bibinfo{author}{\bibfnamefont{L.~A.} \bibnamefont{Pando~Zayas}} \bibnamefont{and} \bibinfo{author}{\bibfnamefont{N.}~\bibnamefont{Quiroz}}, \bibinfo{journal}{JHEP} \textbf{\bibinfo{volume}{11}}, \bibinfo{pages}{023} (\bibinfo{year}{2016}), \eprint{1605.08666}.

\bibitem[{\citenamefont{Gadelha et~al.}(2006)\citenamefont{Gadelha, Marchioro, and Nedel}}]{GMN}
\bibinfo{author}{\bibfnamefont{A.}~\bibnamefont{Gadelha}}, \bibinfo{author}{\bibfnamefont{D.~Z.} \bibnamefont{Marchioro}}, \bibnamefont{and} \bibinfo{author}{\bibfnamefont{D.~L.} \bibnamefont{Nedel}}, \bibinfo{journal}{Phys. Lett. B} \textbf{\bibinfo{volume}{639}}, \bibinfo{pages}{383} (\bibinfo{year}{2006}), \eprint{hep-th/0605237}.

\bibitem[{\citenamefont{Marchioro and Nedel}(2020)}]{Marchioro:2020qub}
\bibinfo{author}{\bibfnamefont{D.~F.~Z.} \bibnamefont{Marchioro}} \bibnamefont{and} \bibinfo{author}{\bibfnamefont{D.~L.} \bibnamefont{Nedel}}, \bibinfo{journal}{JHEP} \textbf{\bibinfo{volume}{07}}, \bibinfo{pages}{102} (\bibinfo{year}{2020}), \eprint{2005.09040}.

\bibitem[{\citenamefont{Papadopoulos et~al.}(2003)\citenamefont{Papadopoulos, Russo, and Tseytlin}}]{PRT}
\bibinfo{author}{\bibfnamefont{G.}~\bibnamefont{Papadopoulos}}, \bibinfo{author}{\bibfnamefont{J.}~\bibnamefont{Russo}}, \bibnamefont{and} \bibinfo{author}{\bibfnamefont{A.~A.} \bibnamefont{Tseytlin}}, \bibinfo{journal}{Class. Quant. Grav.} \textbf{\bibinfo{volume}{20}}, \bibinfo{pages}{969} (\bibinfo{year}{2003}), \eprint{hep-th/0211289}.

\bibitem[{\citenamefont{Marchioro and Nedel}(2008)}]{Marchioro:2007ch}
\bibinfo{author}{\bibfnamefont{D.~Z.} \bibnamefont{Marchioro}} \bibnamefont{and} \bibinfo{author}{\bibfnamefont{D.~L.} \bibnamefont{Nedel}}, \bibinfo{journal}{Eur. Phys. J. C} \textbf{\bibinfo{volume}{55}}, \bibinfo{pages}{343} (\bibinfo{year}{2008}), \eprint{0711.0556}.

\bibitem[{\citenamefont{Chen et~al.}(2006)\citenamefont{Chen, He, and Zhang}}]{Bin}
\bibinfo{author}{\bibfnamefont{B.}~\bibnamefont{Chen}}, \bibinfo{author}{\bibfnamefont{Y.-l.} \bibnamefont{He}}, \bibnamefont{and} \bibinfo{author}{\bibfnamefont{P.}~\bibnamefont{Zhang}}, \bibinfo{journal}{Nucl. Phys.} \textbf{\bibinfo{volume}{B741}}, \bibinfo{pages}{269} (\bibinfo{year}{2006}), \eprint{hep-th/0509113}.

\bibitem[{\citenamefont{Hashizume and Suzuki}(2013)}]{Hashizume_2013}
\bibinfo{author}{\bibfnamefont{Y.}~\bibnamefont{Hashizume}} \bibnamefont{and} \bibinfo{author}{\bibfnamefont{M.}~\bibnamefont{Suzuki}}, \bibinfo{journal}{Physica A: Statistical Mechanics and its Applications} \textbf{\bibinfo{volume}{392}}, \bibinfo{pages}{3518–3530} (\bibinfo{year}{2013}), ISSN \bibinfo{issn}{0378-4371}, \urlprefix\url{http://dx.doi.org/10.1016/j.physa.2013.04.022}.

\bibitem[{\citenamefont{Takesaki}(1970)}]{Takesaki:1970aki}
\bibinfo{author}{\bibfnamefont{M.}~\bibnamefont{Takesaki}}, \emph{\bibinfo{title}{{Tomita's Theory of Modular Hilbert Algebras and its Applications}}}, Lecture Notes in Mathematics (\bibinfo{publisher}{Springer-Verlag}, \bibinfo{year}{1970}).

\bibitem[{\citenamefont{Landsman and van Weert}(1987)}]{lands}
\bibinfo{author}{\bibfnamefont{N.~P.} \bibnamefont{Landsman}} \bibnamefont{and} \bibinfo{author}{\bibfnamefont{C.~G.} \bibnamefont{van Weert}}, \bibinfo{journal}{Phys. Rept.} \textbf{\bibinfo{volume}{145}}, \bibinfo{pages}{141} (\bibinfo{year}{1987}).

\bibitem[{\citenamefont{Haag et~al.}(1967)\citenamefont{Haag, Hugenholtz, and Winnink}}]{Haag:1967sg}
\bibinfo{author}{\bibfnamefont{R.}~\bibnamefont{Haag}}, \bibinfo{author}{\bibfnamefont{N.~M.} \bibnamefont{Hugenholtz}}, \bibnamefont{and} \bibinfo{author}{\bibfnamefont{M.}~\bibnamefont{Winnink}}, \bibinfo{journal}{Commun. Math. Phys.} \textbf{\bibinfo{volume}{5}}, \bibinfo{pages}{215} (\bibinfo{year}{1967}).

\bibitem[{\citenamefont{Dias et~al.}(2021{\natexlab{b}})\citenamefont{Dias, Nedel, and Senise~Jr.}}]{Dias:2019ezx}
\bibinfo{author}{\bibfnamefont{M.}~\bibnamefont{Dias}}, \bibinfo{author}{\bibfnamefont{D.~L.} \bibnamefont{Nedel}}, \bibnamefont{and} \bibinfo{author}{\bibfnamefont{C.~R.} \bibnamefont{Senise~Jr.}}, \bibinfo{journal}{Braz. J. Phys.} \textbf{\bibinfo{volume}{51}}, \bibinfo{pages}{1145} (\bibinfo{year}{2021}{\natexlab{b}}), \eprint{1910.11427}.

\bibitem[{\citenamefont{Botta~Cantcheff et~al.}(2018)\citenamefont{Botta~Cantcheff, Gadelha, Marchioro, and Nedel}}]{BottaCantcheff:2017kys}
\bibinfo{author}{\bibfnamefont{M.}~\bibnamefont{Botta~Cantcheff}}, \bibinfo{author}{\bibfnamefont{A.~L.} \bibnamefont{Gadelha}}, \bibinfo{author}{\bibfnamefont{D.~F.~Z.} \bibnamefont{Marchioro}}, \bibnamefont{and} \bibinfo{author}{\bibfnamefont{D.~L.} \bibnamefont{Nedel}}, \bibinfo{journal}{Eur. Phys. J. C} \textbf{\bibinfo{volume}{78}}, \bibinfo{pages}{105} (\bibinfo{year}{2018}), \eprint{1702.02069}.

\bibitem[{\citenamefont{Ojima}(1981{\natexlab{a}})}]{ojima81}
\bibinfo{author}{\bibfnamefont{I.}~\bibnamefont{Ojima}}, \bibinfo{journal}{Annals Phys.} \textbf{\bibinfo{volume}{137}}, \bibinfo{pages}{1} (\bibinfo{year}{1981}{\natexlab{a}}).

\bibitem[{\citenamefont{Maldacena}(2003)}]{Maldacena:2001kr}
\bibinfo{author}{\bibfnamefont{J.~M.} \bibnamefont{Maldacena}}, \bibinfo{journal}{JHEP} \textbf{\bibinfo{volume}{04}}, \bibinfo{pages}{021} (\bibinfo{year}{2003}), \eprint{hep-th/0106112}.

\bibitem[{\citenamefont{Gross et~al.}(1982)\citenamefont{Gross, Perry, and Yaffe}}]{Gross:1982cv}
\bibinfo{author}{\bibfnamefont{D.~J.} \bibnamefont{Gross}}, \bibinfo{author}{\bibfnamefont{M.~J.} \bibnamefont{Perry}}, \bibnamefont{and} \bibinfo{author}{\bibfnamefont{L.~G.} \bibnamefont{Yaffe}}, \bibinfo{journal}{Phys. Rev. D} \textbf{\bibinfo{volume}{25}}, \bibinfo{pages}{330} (\bibinfo{year}{1982}).

\bibitem[{\citenamefont{Barbon and Rabinovici}(2002)}]{Barbon:2001di}
\bibinfo{author}{\bibfnamefont{J.~L.~F.} \bibnamefont{Barbon}} \bibnamefont{and} \bibinfo{author}{\bibfnamefont{E.}~\bibnamefont{Rabinovici}}, \bibinfo{journal}{JHEP} \textbf{\bibinfo{volume}{03}}, \bibinfo{pages}{057} (\bibinfo{year}{2002}), \eprint{hep-th/0112173}.

\bibitem[{\citenamefont{Brustein and Zigdon}(2022)}]{Brustein:2022uft}
\bibinfo{author}{\bibfnamefont{R.}~\bibnamefont{Brustein}} \bibnamefont{and} \bibinfo{author}{\bibfnamefont{Y.}~\bibnamefont{Zigdon}}, \bibinfo{journal}{JHEP} \textbf{\bibinfo{volume}{05}}, \bibinfo{pages}{031} (\bibinfo{year}{2022}), \eprint{2201.03541}.

\bibitem[{\citenamefont{Salomonson and Skagerstam}(1989)}]{Salomonson:1988ac}
\bibinfo{author}{\bibfnamefont{P.}~\bibnamefont{Salomonson}} \bibnamefont{and} \bibinfo{author}{\bibfnamefont{B.~S.} \bibnamefont{Skagerstam}}, \bibinfo{journal}{Physica A} \textbf{\bibinfo{volume}{158}}, \bibinfo{pages}{499} (\bibinfo{year}{1989}).

\bibitem[{\citenamefont{Salomonson and Skagerstam}(1986)}]{Salomonson:1985eq}
\bibinfo{author}{\bibfnamefont{P.}~\bibnamefont{Salomonson}} \bibnamefont{and} \bibinfo{author}{\bibfnamefont{B.-S.} \bibnamefont{Skagerstam}}, \bibinfo{journal}{Nucl. Phys. B} \textbf{\bibinfo{volume}{268}}, \bibinfo{pages}{349} (\bibinfo{year}{1986}).

\bibitem[{\citenamefont{Cantcheff et~al.}(2012)\citenamefont{Cantcheff, Gadelha, Marchioro, and Nedel}}]{adsnos}
\bibinfo{author}{\bibfnamefont{M.~B.} \bibnamefont{Cantcheff}}, \bibinfo{author}{\bibfnamefont{A.~L.} \bibnamefont{Gadelha}}, \bibinfo{author}{\bibfnamefont{D.~F.~Z.} \bibnamefont{Marchioro}}, \bibnamefont{and} \bibinfo{author}{\bibfnamefont{D.~L.} \bibnamefont{Nedel}}, \bibinfo{journal}{Phys. Rev.} \textbf{\bibinfo{volume}{D86}}, \bibinfo{pages}{086006} (\bibinfo{year}{2012}), \eprint{1205.3438}.

\bibitem[{\citenamefont{Nedel et~al.}(2004)\citenamefont{Nedel, Abdalla, and Gadelha}}]{Nedel:2004gy}
\bibinfo{author}{\bibfnamefont{D.~L.} \bibnamefont{Nedel}}, \bibinfo{author}{\bibfnamefont{M.~C.~B.} \bibnamefont{Abdalla}}, \bibnamefont{and} \bibinfo{author}{\bibfnamefont{A.~L.} \bibnamefont{Gadelha}}, \bibinfo{journal}{Phys. Lett. B} \textbf{\bibinfo{volume}{598}}, \bibinfo{pages}{121} (\bibinfo{year}{2004}), \eprint{hep-th/0405258}.

\bibitem[{\citenamefont{Abdalla et~al.}(2005{\natexlab{a}})\citenamefont{Abdalla, Gadelha, and Nedel}}]{Abdalla:2005qs}
\bibinfo{author}{\bibfnamefont{M.~C.~B.} \bibnamefont{Abdalla}}, \bibinfo{author}{\bibfnamefont{A.~L.} \bibnamefont{Gadelha}}, \bibnamefont{and} \bibinfo{author}{\bibfnamefont{D.~L.} \bibnamefont{Nedel}}, \bibinfo{journal}{JHEP} \textbf{\bibinfo{volume}{10}}, \bibinfo{pages}{063} (\bibinfo{year}{2005}{\natexlab{a}}), \eprint{hep-th/0508195}.

\bibitem[{\citenamefont{Abdalla and Gadelha}(2004)}]{Abdalla:2003xg}
\bibinfo{author}{\bibfnamefont{M.~C.~B.} \bibnamefont{Abdalla}} \bibnamefont{and} \bibinfo{author}{\bibfnamefont{A.~L.} \bibnamefont{Gadelha}}, \bibinfo{journal}{Phys. Lett. A} \textbf{\bibinfo{volume}{322}}, \bibinfo{pages}{31} (\bibinfo{year}{2004}), \eprint{hep-th/0309254}.

\bibitem[{\citenamefont{Green et~al.}(1988)\citenamefont{Green, Schwarz, and Witten}}]{Green:1987sp}
\bibinfo{author}{\bibfnamefont{M.~B.} \bibnamefont{Green}}, \bibinfo{author}{\bibfnamefont{J.~H.} \bibnamefont{Schwarz}}, \bibnamefont{and} \bibinfo{author}{\bibfnamefont{E.}~\bibnamefont{Witten}}, \emph{\bibinfo{title}{{SUPERSTRING THEORY. VOL. 1: INTRODUCTION}}}, Cambridge Monographs on Mathematical Physics (\bibinfo{year}{1988}), ISBN \bibinfo{isbn}{978-0-521-35752-4}.

\bibitem[{\citenamefont{Ojima}(1981{\natexlab{b}})}]{Ojima:1981ma}
\bibinfo{author}{\bibfnamefont{I.}~\bibnamefont{Ojima}}, \bibinfo{journal}{Annals Phys.} \textbf{\bibinfo{volume}{137}}, \bibinfo{pages}{1} (\bibinfo{year}{1981}{\natexlab{b}}).

\bibitem[{\citenamefont{Abdalla et~al.}(2005{\natexlab{b}})\citenamefont{Abdalla, Gadelha, and Nedel}}]{Abdalla:2004dg}
\bibinfo{author}{\bibfnamefont{M.~C.~B.} \bibnamefont{Abdalla}}, \bibinfo{author}{\bibfnamefont{A.~L.} \bibnamefont{Gadelha}}, \bibnamefont{and} \bibinfo{author}{\bibfnamefont{D.~L.} \bibnamefont{Nedel}}, \bibinfo{journal}{Phys. Lett. B} \textbf{\bibinfo{volume}{613}}, \bibinfo{pages}{213} (\bibinfo{year}{2005}{\natexlab{b}}), \eprint{hep-th/0410068}.

\bibitem[{\citenamefont{Banerjee and Wilkerson}(2017)}]{doi:10.1142/S1793042117501135}
\bibinfo{author}{\bibfnamefont{S.}~\bibnamefont{Banerjee}} \bibnamefont{and} \bibinfo{author}{\bibfnamefont{B.}~\bibnamefont{Wilkerson}}, \bibinfo{journal}{International Journal of Number Theory} \textbf{\bibinfo{volume}{13}}, \bibinfo{pages}{2097} (\bibinfo{year}{2017}).

\bibitem[{\citenamefont{Dimov et~al.}(2017)\citenamefont{Dimov, Mladenov, Rashkov, and Vetsov}}]{Dimov:2017ryz}
\bibinfo{author}{\bibfnamefont{H.}~\bibnamefont{Dimov}}, \bibinfo{author}{\bibfnamefont{S.}~\bibnamefont{Mladenov}}, \bibinfo{author}{\bibfnamefont{R.~C.} \bibnamefont{Rashkov}}, \bibnamefont{and} \bibinfo{author}{\bibfnamefont{T.}~\bibnamefont{Vetsov}}, \bibinfo{journal}{Phys. Rev. D} \textbf{\bibinfo{volume}{96}}, \bibinfo{pages}{126004} (\bibinfo{year}{2017}), \eprint{1705.01873}.

\bibitem[{\citenamefont{Blau et~al.}(2003)\citenamefont{Blau, O'Loughlin, Papadopoulos, and Tseytlin}}]{Blau:2003rt}
\bibinfo{author}{\bibfnamefont{M.}~\bibnamefont{Blau}}, \bibinfo{author}{\bibfnamefont{M.}~\bibnamefont{O'Loughlin}}, \bibinfo{author}{\bibfnamefont{G.}~\bibnamefont{Papadopoulos}}, \bibnamefont{and} \bibinfo{author}{\bibfnamefont{A.~A.} \bibnamefont{Tseytlin}}, \bibinfo{journal}{Nucl. Phys. B} \textbf{\bibinfo{volume}{673}}, \bibinfo{pages}{57} (\bibinfo{year}{2003}), \eprint{hep-th/0304198}.

\end{thebibliography}
\end{document}